# Artificial Intelligence for Cybersecurity: Threats, Attacks and Mitigation

Abhilash Chakraborty[1], Anupam Biswas[1], and Ajoy Kumar Khan[2]

[1] Department of Computer Science and Engineering, National Institute of Technology Silchar, India {abhilash21_rs,anupam}@cse.nits.ac.in
[2] Department of Computer Engineering, Mizoram University, Aizawl, India ajoyiitg@gmail.com

**Abstract.** With the advent of the digital era, every day-to-day task is automated due to technological advances. However, technology has yet to provide people with enough tools and safeguards. As the internet connects more-and-more devices around the globe, the question of securing the connected devices grows at an even spiral rate. Data thefts, iden-tity thefts, fraudulent transactions, password compromises, and system breaches are becoming regular everyday news. The surging menace of cyber-attacks got a jolt from the recent advancements in Artificial Intelligence. AI is being applied in almost every field of different sciences and engineering. The intervention of AI not only automates a particular task but also improves efficiency by many folds. So it is evident that such a scrumptious spread would be very appetizing to cybercriminals. Thus the conventional cyber threats and attacks are now "intelligent" threats. This article discusses cybersecurity and cyber threats along with both conventional and intelligent ways of defense against cyber-attacks. Furthermore finally, end the discussion with the potential prospects of the future of AI in cybersecurity.

Keywords: Cybersecurity · Cyber-attacks · DDoS · Man-in-the Middle · Intrusion Detection · Artificial Intelligence.

## 1 Introduction

Nowadays, it is hard to find a company, institution, or family that is not using the internet and technology. We, as humans, find ourselves overwhelmed with the number of digital devices and apps that we use on a day-to-day basis. Many of us cannot even control our technology usage the way we want to. Some peo-ple may be addicted to the internet and cannot stop using it, while others may not use it enough to keep up with the rapid changes in technology. Some might spend hours on their phones or computers rather than interacting with others. The internet is a powerful tool, and we cannot resist the urge to use it for ev-erything. There is no doubt that technology has increased our productivity and efficiency in several ways. However, we have to consider the effects it has on our personal and social lives and our mental and physical well-being. With the bright side of the proliferation of Internet-connected devices, a darker side of cyber-crimes has loomed shadow over our lives. The ever-lurking threat of losing our privacy in this open and connected world has raised many questions. This sudden increase in interconnectedness has made us more efficient and made us vulnerable to the dangers of cyber-crime simultaneously. We must be conscious of our connection to cyberspace. The internet has made us conquerors and pris-oners simultaneously. The internet has given people the ability to connect in unprecedented ways. However, this has also created vulnerabilities that can be exploited by cybercriminals.

Cyber-crime is a type of crime that uses digital media to commit fraud, steal data, or cause damage. Cyber-crime is an umbrella term that encompasses all forms of cyber-related crimes; in essence, illegal activities that are initiated us-ing computers, such as hacking, phishing, malware distribution, online stalking, and identity theft, among many others. Cyber-crime is one of the most lucrative crimes of the modern age. Every year, cybercriminals make billions of dollars in profits by stealing data, corrupting data, and compromising critical infrastruc-ture. Much like any other crime, cybercrime has evolved dramatically over time and will continue to do so.

The report from "The World Economic Forum" states that if we take a holistic approach to cybersecurity, there will be better protection for business, society, governments, and individuals alike [2]. It also states that the gap be-tween our current state of readiness and what is required to protect cyberspace is vast. There is an urgent need to close this gap before we reach a point of no return. Cyber Security has become a critical aspect for every infrastructure not to get compromised by any kind of attack originating from outside sources such as viruses, malware, or hacking attempts. However, most security breaches or cybercrimes are due to human error, not through an external threat. The responsibility of safeguarding the critical infrastructure and protecting people from cybercrime is a shared responsibility for all.

Although viruses and malware were a concern almost from the beginning of computing, awareness about how important data security has only become apparent as the internet grew in popularity. There has been a recent surge in hacking, and it's exposing these "cybercriminals" to more possibilities on the internet. That being said, there are many threats that arise from this exposure. Hackers can cause things like downtime (e.g., by crashing your website), steal data from our computers/servers, or even commit fraudulent transactions. It has now become a separate branch of criminal investigations called cybercrime.

This is because of the extensive global internet penetration of about 5 billion users, which is approximately 63% of the world's population, and cybercrime has increased at an exponential rate. Internationally, the calculated damages of cybercrime are roughly about 6 trillion USD by 2021 [1], and has become the third-largest economy in the world if it were measured separately. Cybercrime is expected to cost a total of $10.5 trillion by 2025 [31], up from just $3 trillion in 2015 according to Cybersecurity Ventures. This is arguably the most significant transfer of wealth in human history. It throws off incentives for innovations and will be more profitable than anything we've ever seen. Costs of cybercrime can include damage or destruction of data, hacked data being deleted or restored, money stolen, reduced productivity, intellectual property theft, theft of personal and financial data and embezzlement (taking assets for the use for the credited person who is responsible for the crime) and reputational harm. Creating a more secure system from the start, preventing cyber-crime from happening, and reducing its impact when it does happen has become a multi-disciplinary affair.

Cybercrime is a growing problem, and it is essential to protect ourselves against it. There are many ways to do so, but the most important thing is to be aware of our surroundings and what we do online with our personal information. Just by being aware of our activities and the risk they might incur, we might be able to avoid most of the threats creeping online. There are many ways that one can avail to protect themselves against cybercrime. One way is to ensure the de-vice's safety and security by using antivirus software, internet security software, and firewall software. Another way is by using strong passwords and changing them often. Lastly, keeping the operating system updated with the latest patches and updates. One can also monitor network traffic for vulnerabilities and set up auto-responders to avoid phishing attacks. Furthermore, we should also manage our social media settings and avoid using unsecured Wi-Fi networks in public places. Even just by minimizing how much personal information we share online, we can avoid the risk of being a target of identity theft, cyberstalking, and many more such threats.

In addition to these conventional methods, which are just a stopgap measure at best, the use of AI is an emerging field in the world of cybersecurity. Nowadays, AI is prevalent in almost any and all fields of science, whether from medicine to business or from the military to law enforcement. The use of AI in science is almost ubiquitous. The use of AI in cybercrime is growing at such a rapid rate that it has become one of the significant areas of concern worldwide. AI is a potent tool that is being used to combat many different types of crime. It will be vital for law enforcement agencies worldwide to find new ways to utilize this technology to keep up with the ever-increasing rate of cybercrime. AI is being applied to crime-fighting in a number of different ways. In the case of cybercrime, AI is being used to help identify potential threats, detect patterns that can lead to previous criminal activity, and detect new forms of existing criminal activity. However, AI is also being used as part of a broader research initiative on cybercrime and its perpetrators. Cybercrime data is collected, analyzed, and used to build sophisticated virtual crime scenes that can predict crimes before they happen.

AI can be used to mine data, identify patterns, and predict future events. It can also be used to detect cyber-attacks and prevent them from happening. In the future, AI systems will be able to detect patterns that are not readily apparent to humans, like a possible cyber-attack, by analyzing network traffic and determining if different strings of data are accessed in the same unusual pattern. AI can do many things, and it will continue to evolve and grow to be used in more everyday aspects of our lives.

The entirety of this chapter is divided into a total of four sections. The first section outlines the concept of cybersecurity along with the threats and attack models that hackers commonly use to compromise a computer system. The sec-ond section entails the conventional approaches and methods of mitigating the risks of cyber-attacks. The third section then discusses the AI-based approaches to counter cyber-threats or at least mitigate the risks associated with cyber-attacks. Finally, the fourth and the last section talks about the future scope of AI in cybersecurity.

## 2   Cybersecurity

Cybersecurity is the practice of protecting critical systems and sensitive infor-mation from digital attacks. There are many ways to safeguard data and or-ganizational infrastructure, including intrusion detection, malware protection, strict adherence to sound security practices, and many more. A cyber security threat can be a cyber-attack using malware or ransomware to gain access to data, disrupt digital operations, or damage information. There are all kinds of cyber threats, including corporate spies, hackers, and terrorists [28]. In fig. 1, the taxonomy of cybersecurity is presented. Although all have different reasons for attacking, all should be treated with extreme caution as they pose a risk to an organization's and personal data. The rise of the Internet has brought a new era of cyber security concerns. In addition to the threat of criminal hackers and foreign governments, new challenges are being associated with protecting infor-mation from internal threats, such as data breaches and insider theft. Cyber security is also an essential cross-cutting concern for sensitive infrastructures, critical assets, and sensitive information. This is why there has been a remark-able rise in cyber security professionals and the industry as a whole and why it is becoming increasingly important to ensure that the defense mechanisms against cyber attacks are comprehensive and robust.

Cybersecurity is a broad term encompassing all measures taken in an effort to safeguard an entity from cyber threats, including securing data and mitigating damage from a cyber security incident. The field of Cybersecurity can be broadly classified into five distinct security areas:

– Critical infrastructure security
– Application security
– Network security
– Cloud security and
– Internet of Things (IoT) security.

Cybersecurity is a complex and ever-changing field. It is essential to un-derstand the different types of cyber threats and how they can be mitigated. Cyberattacks are becoming a common occurrence in today's society. However, these attacks can be prevented with the proper security measures.

In this article, we will discuss the different types of cyber-attacks and threats and deep dive into different defense mechanisms, both conventional and AI-based, and learn about the currently available threat mitigating solutions.

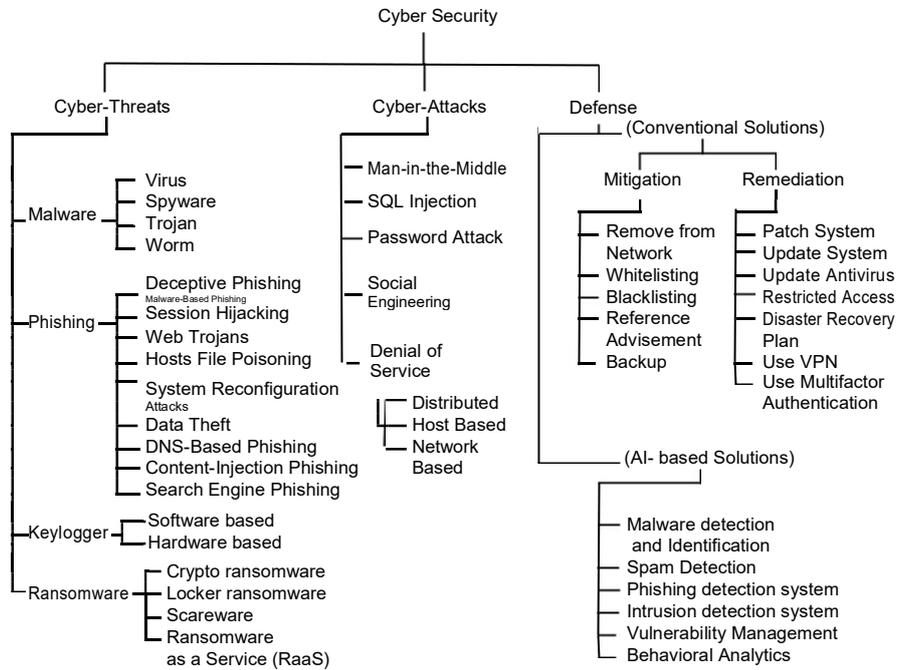

Fig. 1. Taxonomy of Cybersecurity

2.1 Attacks

Distributed Denial of Service or DDoS : It is a form of cyber-attack where the perpetrator uses multiple systems to flood the target with traffic. The goal is to make it difficult for the target to provide service or access their website. The most common type of DDoS attack is a volumetric attack, which floods the target with an overwhelming amount of data. This can be done by using a botnet, a network of computers that have been infected with malware and are controlled by an attacker without their owner's knowledge. Within the volume attacks category, there are flooding and amplification/reflection attacks. In a flooding attack, traffic is sent in the hopes of exhausting bandwidth, processing capacity, or other network resources. Amplification/reflection attacks seek to force victims to spend money by "overloading" their networks with spam traffic or denying access to certain resources using spam-like messages [36] [27].

There are many tools that can be used to launch a DDoS attack, and the most common is the use of a botnet. The attacker does not need to control the botnet, as they may simply rent it from an online service or purchase it from someone else. Some examples of these services include Blackhole, Stresser, and NitrousDDoS [37].

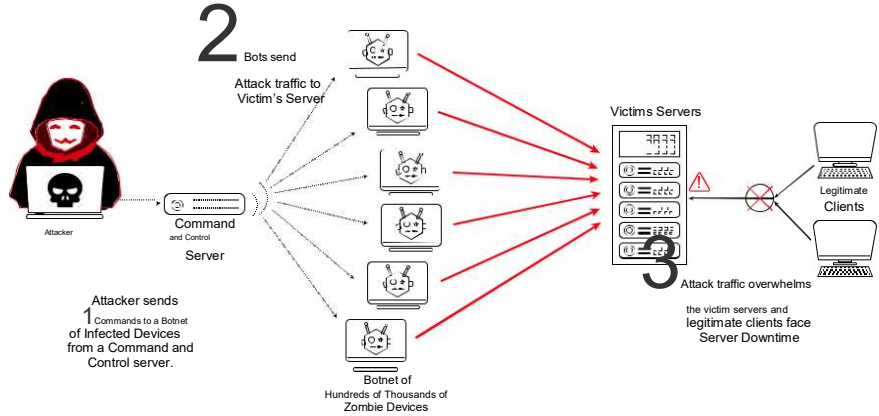

Fig. 2. Distributed Denial of Services Attack scenario.

Man-in-the-middle attack: A form of cyberattack where the attacker secretly relays and possibly alters the communication between two parties who believe they are directly communicating with each other is called a Man in the Middle attack. The attacker can read all messages passing between the two systems and can also inject forged messages. The term "man in the middle" comes from an analogy to espionage: one party thinks they are talking directly to another party, while in fact both their messages are being read by an eavesdropper. The man-in-the-middle attack can be made in a number of ways. One way is for the attacker to physically place themselves between the two parties without either party knowing and then relay messages from one party to the other. This could be done, for example, by having access to a telephone company's network and rerouting calls or by being on a public Wifi hotspot. Another way is for an attacker with system admin privileges on a remote computer to use a man-in-the-middle (MITM) exploit to intercept and control traffic between the client and server.

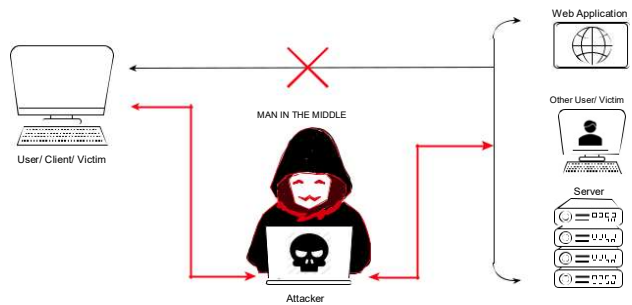

Fig. 3. A simple depiction of Man in the Middle Attacks

It is often referred to as "man-in-the-browser" because it uses vulnerabili-ties in a browser or other software to accomplish the attack. The attack can use social engineering, where the attacker tricks the user into accepting an un-safe connection over HTTPS or other secure protocols, or by exploiting known vulnerabilities in the software, such as Cross-Site Scripting. It is also used to refer to attacks where an unauthorized person gains access to a computer run-ning a browser, uses the browser's interface via webcam and microphone, and recording capabilities, in order to spy on the user. The attacker can then use this information for blackmail or other nefarious purposes.

SQL injection: This type of cyber-attack exploits the security vulnerability in the database. It is a type of code injection technique that can be used to attack data-driven applications. The SQL injection attack is one of the most common and dangerous types of cyber-attacks. It can be used to steal sensitive information from databases, modify or delete data, and disrupt service. SQL injection uses the dynamic nature of SQL (structured query language) to cir-cumvent input validation and access data that is otherwise inaccessible. SQL injection usually involves the use of malformed or clearly erroneous input in a SQL command. For example, capturing the password after inserting into an ac-count table may involve sending "select password from users" instead of "insert into users values (username, password)". This causes the database to execute the query and return the username, resulting in a list of user names from which the attacker can extract a database user's password. SQL injection may also be used to change data in a table without proper authorization. This may result in loss of confidentiality and availability for that particular table or other tables that reference it.

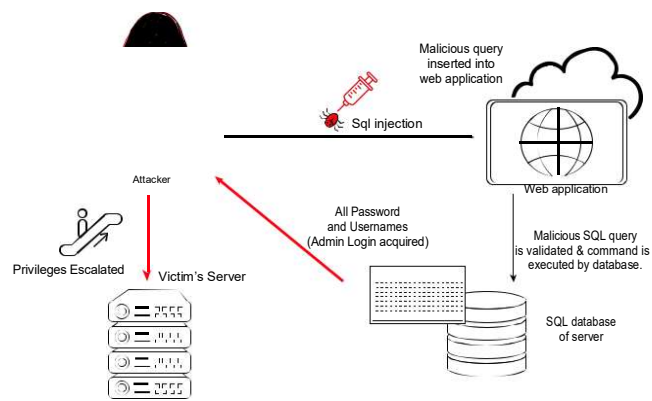

Fig. 4. A typical illustration of SQL injection attack.

A password attack: A method of cyber-attack where the attacker attempts to guess passwords or steal them outright, usually by hacking into a computer system or network is referred to as a Password attack. The attacker may use a man-in-the-middle attack to intercept the victim's password and then use it to access their account. It is notoriously difficult to crack a password, but with advanced cracking programs and tactics, hackers can eventually breakthrough. There are three types of password attacks: brute force, dictionary, and keylogging. Knowing these can help reduce your risk of attack. A brute force attack is a series of attempts at guessing passwords until the attacker penetrates the system. Dictionary attacks involve trying different combinations of dictionary words before stumbling upon the password. Keylogging is a method that records keystrokes to extract sensitive data like login credentials for use in a password recovery attack. This type of attack can be prevented by using two-factor au-thentication or by not clicking on links in emails from unknown sources.

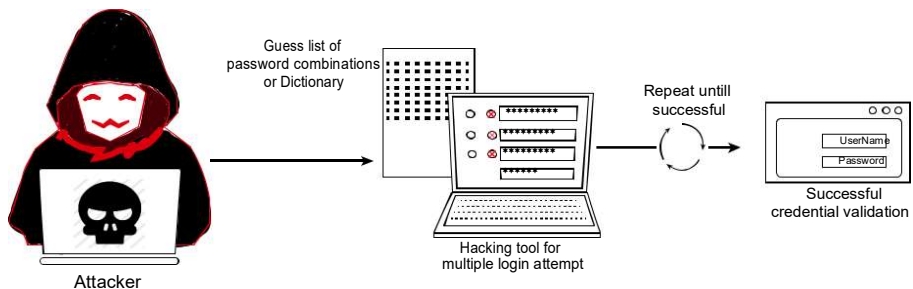

Fig. 5. Depiction of generalised principle of a Password Attack

Attacks on IoT Devices: The Internet of Things (IoT) is a network of physical objects that are embedded with electronics, software, sensors, and connectivity to enable these objects to collect and exchange data. There are multiple IoT net-works in existence, with the most common being the Internet and intranets. The IoT is a new frontier for cybercriminals. They are targeting IoT devices because they are often not appropriately secured. This makes them an easy target for cyber-attacks. As the number of Internet-connected devices continues to grow, it becomes increasingly important to protect these devices from being compro-mised by malicious hackers and cybercriminals. In general, IoT security requires three main components: IoT Device Software Architecture, Trusted Platform Module (TPM), and Security Standards [14].

According to a study published by Cisco Systems, "Cyber Threats to the Internet of Things: Emerging Risks and Tactical Strategies", it has been found that attackers are targeting the growing number of connected devices on the Internet in order to gain access to private information, cause disruption or steal money from users. The study found that the Internet of Things will significantly impact business and consumer perspectives, and it is important for organizations to consider cybersecurity risks. Organizations should review their existing network products and services against these new risks.

With the recent pandemic, the work-from-home culture got a colossal thrust, and it seems like the trend of home-based offices is here to stay. With this precedent, residential areas are now becoming a valuable target for various reasons. Most organizations are well prepared to defend against cyber threats, but that is primarily true from inside the organization's infrastructure. The devices that connect to the organization's servers from the employees' homes become a prime vulnerability point for exploitation by the attackers. The volume of these attacks increased by 35% in the first half of 2020 compared to the second half of 2019, as stated by a report from Microsoft. With the popularity of home-based offices, there is a greater risk of these devices being targeted and exploited by hackers. If a hacker focused on compromising one or more homes in an area could wreak havoc within the organization, then it is highly plausible that they would be able to compromise other homes as well. For example, A de-authentication at-tack on an unsecured wireless network may provide the attacker with a hashed password. Furthermore, this means that the password can ultimately be cracked offline and malicious use of this password is possible.

## 2.2   Threats

Malware: Malwares are a type of malicious software that can be used to steal information, damage or disable computer systems, and gain access to private computer networks. Malware is often disguised as legitimate software or 'adware', or appears as genuine software but performs some hidden function when executed.

Malware can be classified into different categories based on the method of infection:

– Viruses: These are programs that attach themselves to other programs and replicate themselves by infecting other files. They are usually spread through email attachments, downloads from the internet, or by copying infected files onto a CD or DVD.
– Worms: These are self-replicating programs that use a computer's network connections to spread themselves over the internet. They do not need any user intervention in order to spread themselves.
– Trojan horses: These are programs that pretend to be something else, often a valuable item, while performing malicious actions.

Cisco has reported that malware can block key network access, install more harmful software, and transmit data from hard drives. The company is not nam-ing any of the companies involved in the incident, but it does say that Cisco Ta-los cybersecurity researchers observed a type of malware called Responder being used to launch DDoS attacks. The company notes that Responder was created by a hacker named Peter Severa and can be downloaded from underground forums.

Phishing: It is a cyber-threat that is becoming more and more popular. It is a form of social engineering that tricks people into giving up their personal information. The most common form of phishing is when an attacker sends an email to a victim to trick them into clicking on a link or opening an attachment. The link may lead to a fake website that looks like the real one and asks for login credentials. The attachment may be malware that can steal data from the computer or install ransomware. Phishing emails are often designed to look like they come from legitimate companies and organizations to fool people into thinking they are safe and secure.

Phishing scams are becoming more and more sophisticated, using malware-based phishing, keyloggers, session hijacking, web trojans, DNS-based phishing, and host file poisoning to capture victims [35]. In this type of scam, cybercrim-inals send emails that appear to be from trusted companies, requesting personal information like account numbers, passwords, and credit card numbers. The goal is to trick the user into thinking they are interacting with a legitimate company in order to gain access to their accounts or steal their identity.

The most common form of phishing is deceptive phishing. This type of phish-ing uses fake websites to trick people into giving up their personal information, such as passwords, credit cards, and email addresses. There is no solid evidence that most people have any knowledge about and rely on this type of phishing or the effectiveness of these methods. One advantage to using phishing attacks is that geographic boundaries do not restrict them. This is because a person can use a computer anytime and from anywhere. Another benefit to phishing attacks is that they are considered 'low-risk' because the victim does not receive any-thing in return for giving up their personal information. In contrast, if someone were to use social engineering methods, like taking someone else's identity or creating a fake account with a similar name , to send a message with malicious intent, that would be considered cyberstalking.

Cryptojacking: Another form of cybercrime that involves the use of malware to hijack a computer's processing power to mine cryptocurrency is called Cryp-tojacking. The malware is injected into browsers, scripts, and ads, which will invisibly use computing power from the device it has infected to mine cryptocur-rencies. The term "cryptojacking" was coined in 2017 by cybersecurity company McAfee. It is a portmanteau of "cryptocurrency" and "hijacking." Cryptojacking can be done through various methods, including :

– Malware that infects the victim's device and uses its resources to mine cryp-tocurrency.
– Websites that use scripts to mine cryptocurrency on their visitors' devices without their knowledge or consent.
– Mining software running on an infected device.

The increase in cryptocurrency value has led to an increase in coin-mining malware. Many of these infect devices and hijack its resources to mine cryp-tocurrency. There are also websites that use scripts to mine cryptocurrency on their visitors' devices without their knowledge or consent. One of the most common examples is Coinhive, which mines cryptocurrency on a website's visitors' devices. This can result in a decrease in performance and lower battery life for affected devices. Coin-mining malware has also been used to steal cryptocurrency. Attackers have created mining software that steals cryptocurrency, such as Monero, by secretly hijacking devices to use their power. Although the risk of malicious coin-mining malware is high, most of these viruses are easy to prevent by installing the appropriate protection on your devices.

Ransomware: Ransomware attack is a destructive malware that uses encryption to hold data hostage and demands a ransom to decrypt it. It is a type of cyber-attack that has been on the rise in recent years. The term has also been applied to a certain type of malware, dubbed "ransomware-as-a-service," which is a form of ransomware that can be bought and sold on the darknet through so-called "crypto marketplaces".

The ransomware attack can be carried out by an individual or a group of people who are not affiliated with any government or organization. The attackers usually demand payment in Bitcoin, which makes it challenging to trace them. The attack was first observed in May 2017, when an unidentified ransomware program called Petya infected computers worldwide. Ransomware is a type of malicious software that infects vulnerable devices, like a personal computer or server, and restricts or prevents their everyday use until the victim pays ransom to unlock it. Petya uses a stolen NSA-developed malware to spread by exploiting a vulnerability in the Windows operating system and encrypting all of the data on the computer. The software then creates an image with a ransom note that demands $300 worth of bitcoin for the user to regain access to their files. However, unlike other ransomware programs, Petya cannot be detected. The implications of a ransomware attack include:

– Ransomware attacks can disrupt data and operations, slowing business by spreading fear and uncertainty.
– The costs of a ransomware attack are difficult to calculate, but they are likely substantial.
– If a company's information is encrypted, it is more vulnerable to the effects of a future cyberattack.

Keylogger: One of the most commonly used malware to eavesdrop on a user's passwords and other personal information, such as credit card numbers and passwords for email accounts is a Keylogger. If a computer system has a keyboard attached, it is possible for an attacker to use a keylogger to monitor the user's keystrokes through special hardware or software. The goal of this type of attack is usually to steal passwords, personal information, and other sensi-tive data. Keystroke loggers can be inserted in between the keyboard and the computer's central processing unit, which records keystrokes as decoded by a computer's BIOS or operating system. In some cases, the loggers can be placed on the motherboard itself. Keylogging has been used in espionage, with one no-table example being the FBI investigation into John Walker Lindh (the so-called American Taliban). in August 2001. This is done by first installing a keylogger software on the computer and then taking a picture of the victim's keyboard when it is in use. The keylogger software records keys that are pressed, with data about their position on the keyboard and what programs they were used in. The keylogger software runs as a system service and can be configured to start automatically. This allows the software to collect keystrokes in order to create a profile of the user's computer usage patterns and also for the user's personal records. The log files are stored on the hard drive to make subsequent analysis easier and may be saved for future use. The free software is available in a wide variety of languages and can be installed on any computer with a standard browser. It collects keystrokes in two different forms: "scores" generated by the operating system and "patterns" generated by the user's own typing activity. The log files are stored on the hard drive rather than in RAM. The pattern data is encrypted with a key that is specific to the computer and only accessible by the user's password. This means that there isn't an easy way to extract the patterns from a hard drive without a password or breaking into the machine, but if someone could get into your computer, they could read this data in its precise form.

## 2.3   AI as a tool for Cyber-attacks

One of the key factors that have allowed the internet to exist is its decentralized nature. The internet is not owned by any one entity, making it difficult for any one entity to shut down or control it. This unique aspect of the internet partly led to its success, and it allowed for new technologies like AI to become possible. However, with AI becoming more prevalent, the internet can quickly become a very different place. For instance, if AI can control the flow of information on the internet, it could be used to manipulate public opinion (e.g., give people false information that leads to herd mentality) or even cause war. Probably one of the more famous factors that led to AI becoming possible was the Singularity. Singularity is a speculative concept in which technological growth becomes so rapid and complete that it crosses a point of no return, triggering runaway tech-nological change. The result is a "post-human" era in which intelligent machines surpass human intelligence. At the time of this writing, the idea of AI being able to create computer viruses has become quite popular—the problem of how to stop such an AI has yet to be faced.

   AI can be used to create malware that can evade detection by antivirus software. It can also be used to create fake social media profiles and spread mis-information on social media platforms. AI is used by the military and intelligence communities to identify specific objects in a photo or video. The potential to abuse AI goes hand in hand with its potential to make autonomous decisions such as how many people should die based on a predicted crime rate. AI is being used to predict stock market crashes, a 2019 study showed that over 92% of Forex trading was done by AI and not humans [21]. More than 60% of trades over $10M are currently executed using algorithms, and that number is expected to grow significantly over the next four years.

## 3  Conventional Solutions

The most common defense against cyber threats is network security, including firewalls, intrusion detection systems, antivirus software, and encryption technologies. While network security helps, it is not a solution in and of itself. Experts agree that no system is 100% secure because there will always be vulnerabili-ties that attackers can exploit. Network security is one component of an overall cybersecurity strategy. Consecutively Cloud security refers to protecting data stored in cloud computing environments such as Amazon Web Services (AWS) or Microsoft Azure from cybercriminals. For example, data security is often pro-vided by encryption when storing large amounts of customer data in a cloud environment. This can be done using the public key infrastructure or equivalent technology.

A cybersecurity strategy is a collection of techniques, policies, and procedures used to reduce the impact of any security breaches. It includes steps to mitigate risk from threats such as cyber-attacks, data breaches, and malicious software. A cybersecurity strategy includes several components: One component of a cybersecurity strategy is risk assessment which determines the likelihood that an event will occur and the potential consequences of the event. A risk assessment will often consider different types of threats and vulnerabilities in a business. For example, an assessment may consider whether an organization has a website that can be hacked or if there are weaknesses in password protection. Once risks are assessed, then mitigation measures can be developed to reduce the likelihood or impact of cybersecurity events. Mitigation measures are defined as means of reducing the risks of cybersecurity events. The most commonly used mitigation measures are security controls, encryption, and patching.

Among numerous policies is the Zero trust policy, which is beneficial for organizations that want to establish more robust control over different aspects of the company's digital security. It ensures that companies can manage access to sensitive information by looking at resources and prior user history. The Zero trust policy makes it possible to reduce the risk of data breaches and maintain privacy for employees. The zero-trust policy applies to the following:- Applications and data management- Email communication- Mobile applications and apps that the company owns or provides- Cloud computing, infrastructure, or storage services providers or terminals used by the company. The zero-trust policy is beneficial to organizations and employees and to individuals' privacy. It helps individuals build a well-protected digital identity and create opportunities for access when needed. A zero-trust policy is an approach to information security in which an end-user can access any other user's computer or application without any trust assumption. The Zero trust policy also referred to as the No Trust Policy, is an approach to information security in which an end-user can access any other user's computer or application without creating a trust relationship. This is done by establishing a "penetration of trusted computing" model that requires all users and devices to meet certain requirements before connecting to network resources. The term is also used in understanding cybersecurity: a zero-trust approach has been used to create the concept of an Internet of Things (IoT) that relies on trust-less computing protocols.

Here is a list of strategies undertaken to defend against the cyber threats as far as conventional measures are considered:

– Using Firewall and antivirus programs:

  A firewall is software or hardware that separates the computer system from the internet to prevent it from being infected with viruses and other malicious software. Firewalls act as a buffer between the outside world and the network and give organizations greater control over incoming and outgoing traffic. Similarly, antivirus software, when run, detects and blocks any troubling threats. This process often includes scanning the device and/or network for any possible malware that might be present and then removing it. As one might expect, modern antivirus software can assess computers from two different directions. On the simple side is a system scan, which looks through files and prevents any potential harmful threats from causing further damage to the system. It does not get rid of anything, but it does make sure the system does not have malware. The other method is a comprehensive scan. This looks through every file and erases any possible threats, including viruses and malware. On the downside, it requires quite a bit of time to complete because it needs to look at all of the files and erase them if found malicious, but it prevents any potential for future harm.

– Using Secure browser extensions

  A secure browser extension helps stay safe on the internet by blocking phish-ing and malicious websites that try to steal personal information. It protects against the risk of malware as well as adware too. For instance, to improve on-line privacy, the "HTTPS Everywhere" Firefox extension forces your browser to use an SSL-encrypted version of a website when it is available. "Privacy Badger" automatically stops third-party websites from following your activ-ity on the web. "AdBlock Plus" helps remove advertisements while browsing.

– Using VPNs

  A VPN is the most effective way to protect data from cyberattacks. It en-crypts data and routes it through an encrypted tunnel so that hackers cannot access it. Some VPNs offer a kill switch that stops all internet traffic when the VPN connection is lost. This will prevent any data from leaking out through peer-to-peer connections, which could be helpful in certain situations. The best part of using a VPN is that the data is locked away and secured while traveling between servers. This makes accessing browsing history and data harder for anyone, including the ISPs and hackers. The final destination of the outgoing traffic stays a secret as well. Moreover, by connecting to the VPN, a computer or a device's IP address gets "hidden" behind the one that is given by the VPN server. This is helpful for anonymously browsing online.

– Using Strong and unique passwords

The use of strong and unique passwords are essential because they make it harder for hackers to guess what password used on different websites or apps are. Using a strong password includes uppercase and lowercase letters, numbers, and symbols in a unique but easy-to-remember way. Furthermore, a password manager can organize passwords into categories, so we will not forget which websites or apps require what kind of password.

– Using Security paths and updates

  Security updates are essential because they keep computers up-to-date. It is vital to keep software updated for the sake of the device and everyone else on the network, even if it isn't fun. Once a security update has been released, attackers will try and exploit that software and those who do not use it.

## 4   Intervention of AI

The use of AI in cyberattacks is a new and emerging trend. It is not yet clear how this will affect the future of cybercrime. There are several different AI and machine learning techniques used in cybersecurity. The most common ones in-clude strategies that use AI to identify and monitor malicious activities, detect cyberthreats, and protect an organization's networks. For instance, a malware analyst can use machine learning algorithms to train an AI system on how to detect malicious files or identify compromised PCs. An AI system can also monitor the behavior of an individual or group, such as detecting changes in activity on social media or analyzing the traffic patterns of employees to identify those who might be up to something unusual. When integrating AI into cybersecurity, the key challenges for organizations are how to design and manage data that is available across multiple systems and how to structure data to make it accessi-ble for cognitive applications that can incorporate human supervision. Artificial intelligence has permeated many aspects of our professional and personal lives.

### 4.1   Recent Trends

Along with this trend, cybersecurity is also increasingly adopting cognitive tech-nologies. AI-powered cognitive technologies are an essential part of a holistic approach to cybersecurity in which the human element guides the process and plays a pivotal role. In general, cyber defense is a constantly shifting space where the nature of security threats changes with each new development. Cybersecurity professionals who can adopt successful cognitive technologies and guide their hu-man element on a holistic approach will be more successful in defending against cyber-attacks. The industry has also embraced certain trends that have been years in the making, such as blockchain technology's role as an enabler for cyber defense and the increased need for artificial intelligence in cybersecurity. The report predicts that the IT security workforce size will grow as a result of these shifts. Cybersecurity is a vital component of any business and can be challenging to quantify. In its report, Cybersecurity Trends to Look Out For in 2019, Cy-berVance discusses the importance of examining cybersecurity strategy trends, particularly how organizations can adopt new technology and protect against cyber-attacks.

In the past, cybersecurity professionals focused primarily on monitoring threats and defending against them. Now they are more concerned with risk assessment and mitigation, which allows them to avoid exploits that could cause harm. As a result, the most critical question to ask is, "What is the risk of this type of exploit?" As we can see, there are many other changes to cybersecurity pro-fessionals. They now focus on mitigating risks and assessing probability rather than monitoring threats. These changes create an entirely new world for those looking to enter the field.

A broader classification of the AI techniques used for detecting and mitigating cyberthreats include: Expert Systems and Intelligent agents.

- Expert Systems

  Expert Systems are a type of computer device that provides the decision-making power of a person. Knowledge-based systems are made up of two sub-systems, namely the Knowledge Base and the Inference Engine. The Knowledge Base stores the information and is linked to the Inference En-gine, which interprets it or draws an inference from the available information to make decisions. Knowledge-Based Systems can make predictions and judgments based on the information in the knowledge base. They may be used in tasks such as medical diagnosis, stock trading, or even prognostication of the future. The knowledge-based system is a computer system that combines a computational engine (Inference Engine) and data storage in order to make predictions about unknown variables based on given known variables. Some examples of such systems are Weather Channel, Google search, Alexa or Siri. Knowledge-based systems take an existing body of knowledge and use it to create predictive models that are used in particular scenarios.

- Intelligent Agents

  An intelligent agent is a software that exists in an environment that is not controlled by anyone externally. It can respond to fluctuations in its sur-roundings and continuously pursue its goals over time. They always have multiple ways of achieving those goals. An intelligent agent can be designed to learn all possible actions and then select the best option for accomplishing its goal. Intelligent agents are those that have the ability to learn and adapt to their environment.

Artificial intelligence can be used to detect and stop cyber-attacks by mim-icking human intelligence. It can detect behavior patterns to identify potential threat signals that indicate a potential attack. Machine learning can be used in cyber security to detect and prevent targeted attacks on industrial control sys-tems. Machine learning models can be trained to identify anomalous behavior that matches a targeted attack, thus allowing a cyber security system to block the attack before it is executed automatically. Intrusion detection technologies can be improved by incorporating machine learning techniques into them and using neural networks to detect anomalous behavior in traffic.

Machine learning is used in cyber security to help detect and prevent tar-geted attacks on industrial control systems. Machine learning models can be trained to identify anomalous behavior that matches a targeted attack, thus allowing a cyber security system to block the attack before it is executed au-tomatically. Anomaly detection technologies can be improved by incorporating machine learning as an additional feature of the anomaly detection system. Ma-chine learning can be applied to detect anomalous behavior based on data or anomalous machine-learning models that learn from data. Anomaly detection systems in networks can use machine learning as a metric to determine if anoma-lous activity is present in network traffic and then take actions such as filtering out the traffic in question or even taking further action.A method of anomaly detection consists of four components: input, training data, model parameters, and output. The input is the sequence of observations for which an anomalous event is predicted. This can be a number of characteristics such as TCP port numbers, HTTP header fields, and IP addresses at a company's edge routers. The training data is a collection of sequences of observations that the system has been annotated with. These sequences likely contain anomalous events. The model parameters define the training algorithm and include: normalization pa-rameters, anomaly detection sensitivity, and detection threshold. These are used to measure how well an anomaly detector can identify an event and whether it is in a state of overconfidence or under-confidence. The model parameters also define how the anomaly detector reacts when detecting a false positive. Finally, the outputs of an anomaly detector include confidence and hypotheses. Confi-dence measures the likelihood that an anomalous event is occurring in a given sequence. Hypotheses are possible causes for the event with which an anomaly detector can work to identify a pattern or set of patterns in the data.

### 4.2    AI based mitigation of Cyberthreats

Malware detection and identification: Artificial intelligence being used for malware detection and identification is still in its infancy, but it has the potential to revolutionize the way we deal with cybercrime. AI can help identify malicious files before they reach the end-user and, by doing so, can provide significant security benefits. Many different AI/ML approaches have been used to detect malware, some more successful than others [19]. One approach uses machine learning and data mining to look for malware source code repositories using a technique called "SourceFinder" and analyses them based on characteristics and properties [30]. Another way uses machine learning to look for a particular strings within files that could indicate the presence of malware or malicious code and also classify them [34]. Another approach is to use AI/ML to detect patterns in binary executable files and determine if they are malicious [32] [32]. Another method for detecting and stopping malware is by utilizing visual binary patterns identified in the code and a type of self-organizing network that adapts over time [5].

Systems typically use heuristics - the process of looking for patterns in data - to find malware by crawling through huge data sets and looking for suspicious files that might be indicators of the presence of malware. In other words, systems might use hash values to detect files that have been recently modified, or they might use file size metrics over a period of time to determine that a new file appears to be too large. An antivirus (AV) program is designed to use heuristics and can employ techniques such as pattern matching, statistical analyses, and emulation of known malware signatures.

Coull et al. [9] presented Byte-activation analysis which is a type of neural network where the response to an input is mapped to the activation of each byte. On the other hand the FireEye [20] used the Convolution neural network and here a total of three networks were trained with different parameters, such as training set sizes and dropout settings. Dropout may have been turned off, or on during the process. Training the networks with a particular dataset of 7 million files was done to find the specific parameters. Adversarial Examples are inputs with small, imperceptible variations which cause neural networks to misreport it. Neural networks designed to detect malware automatically may not be immune to these types of attack and is the focus of the work by Demetrio et al. [10]. The purpose of this network is to analyze the structure to create adversarial examples that are misclassified and attempt to challenge it. A similar approach was tried before, but the results found in this study are different from those presented by

[22] and [23].Moreover Bose et al. [7] proves that their technique can provide improved insight into each classification from [9] [10], while also exploring new solution areas to look for a better performance. The architecture includes two filters A and B wherein Filter A is designed to detect goodware content, while Filter B is designed to find specific parts in a file which makes it malicious.

A code obfuscation technique is a challenge for signature based techniques used by advanced malware to evade anti-malware tools. To tackle this, Sharma et al. [33] discusses an approach that they used to improve the accuracy of de-tection of unknown advanced malware and proposed a new method that uses Fisher Score for the feature selection and five classifiers to uncover the unknown malicious. A few other notable works in the same genre are as follows: Chowd-hury et al. [8] have published a paper on classifying and detecting malware using data mining and machine learning. They use these methods to classify malicious websites that could potentially lead to infections. Hashemi et al. [18] used KNN and SVMs strategies to detect unknown malware on the data. Malware detection in android devices is done in [26], which used a deep CNN to identify malware in android devices, and [40], where a novel ML approach called rotation forest was used. Ye et al. [39] presents a new deep learning model, which establishes the SAE model for intelligent malware detection that's built on analyzing Win-dows API calls. The experimental results show that this method can offer a lot more than traditional shallow learning in malware detection and improve overall performance.

Spam Detection: AI can be used to detect spam by analyzing the content of the message and looking for patterns that are typical of spam. This is done by using machine learning algorithms that have been trained on a large number of examples of spam and non-spam messages. In the process of spam detection, AI will often replace a human role and be able to detect that messages are spam without requiring any human intervention. The automated system can also flag messages as potentially being spam and require human review before being classified as such. As an example, Feng et al. [13] presented a system built to filter out some spam emails. It combined a support vector machine and a naive Bayes algorithm.

On the other hand, AI-based spam detection in online review verification is also popular; for instance, A new unsupervised text mining model by Lau et al. [38] was developed in an effort to explore the possibility of detecting false reviews. This method was trained on a semantic language model to identify duplicates in reviews and then compared to supervised learning methods, which have already been successful in the review industry. The dataset was trained with a high-order concept of association. The results were interpreted to extract context-sensitive concept association knowledge among the reviewers and posts.

The Phishing detection system: An artificial intelligence based system that can detect phishing emails by analyzing the content of the email and comparing it with a database of known phishing emails [4] [29]. The system can also detect if the sender is spoofing another person's identity. The phishing detection system can also be used with voice, video, and image messages. The system activates when a user receives a suspicious email or when they send an email containing personal information. Some of the features of the current phishing detection system are:

– Automatically detects email phishing scams
– Stores emails with malicious content in a quarantine folder
– Triggers user notifications when the system detects a new virus in an email.
– Detailed logs of all email activity
– Detects emails that contain phishing links
– Automatically generates a report of every detected email.

The goal of the phishing detection system is to automatically detect and report emails that contain phishing links. For instance, Feng et al. [12] utilized a neural network to detect phishing websites by using the Monte Carlo algo-rithm and risk minimization approach. Another approach by Mahajanet al. [25] proposed a system in Phishing Website Detection using Machine Learning Al-gorithms, which would keep track of various features of legitimate and phishing Uniform Resource Locators (URLs). They deal with machine learning algorithms to detect phishing URLs and use ML techniques to overcome the disadvantages of blacklist and heuristic-based methods, which cannot detect phishing attacks.

Intrusion detection system: The use of AI for intrusion detection systems is a new and emerging field. It is a branch of computer science that deals with developing intelligent systems to detect, classify, and respond to cyber-attacks.

They are designed to identify malicious behavior and stop it before it causes any damage. It can be implemented as a standalone system or as an add-on module to other security software such as antivirus programs. The intrusion detection system is usually configured with a set of rules that define what constitutes an attack, such as the use of certain words in the subject line of an email message or the sending of too many messages in a given period of time. The IDS then compares each packet of data against these rules and takes action if there is a match. The intrusion detection system is usually configured to generate alerts when it detects an event that might indicate an attack or intrusion attempt. IDS responses can be categorized into two main types: An active defense is one in which an IDS initiates a response. For example, it might issue an alert to on-duty personnel about a potential intrusion attempt. An active defense is the closest thing to "real-time" defense because the system initiates an action at the moment of detection rather than waiting for a report from another system. A passive defense is one in which the IDS only responds after receiving and processing information that an intrusion attempt has taken place. An example of this type of response would be when a system that is already installed on-site monitors for changes in network traffic around its perimeter and then initiates a report about these changes to a central monitoring hub and also stores the data in an analytics database.

The goal of an AI-based algorithm is to optimize certain features and improve its classifiers so that it may be able to reduce the number of false alarms that come up while trying to identify an intruder. It can also help to pre-empt and address potential security risks in a company's environment from the moment they are identified. A combination of SVM and a modified k-means was used by Al-Yaseen et al. [3] to create an intrusion detection system. On the other hand Hamamoto et al. [16] used fuzzy logic along with genetic algorithm to detect occurrence any anomaly in network. This was to predict a network's traffic for a given time interval.

A detailed survey of intrusion detection efforts in the last few decades is given in [17]; with many works listed, they conclude that Hybrid Machine Learning techniques have been used widely. Barbara et al. [6] proposed the hybrid Audit Data Analysis and Mining architecture, where the anomaly detection is followed by misuse detection. Farid et al. [11] improved anomaly intrusion detection us-ing the Self Adaptive Bayesian Algorithm, which is designed to be used in large amounts of data. Another approach integrated Correlation-Based Feature Selec-tion to select the best feature set. Resulting in improvement of the detection rate of the reduced data-set, as it selected the best feature set and removed unimportant data-sets [15]. Chowdhury et al. [24] proposed a new technique to reduce the dimensionality of data. Instead of using a traditional neural network, they use a triangular approach to calculate and visualize data.

## 5  Conclusion

AI can detect and stop cyber threats in real-time with limited resources. The constantly evolving nature of cyber-attacks means that humans shall struggle to keep up with the intel. However, using machine learning, AI can chomp down data for quick analysis and provide excellent security coverage without taking much time or energy away from the existing tasks. Machine learning allows Human analysts to focus on interpreting the results from deep analysis and devising novel techniques for fighting cyber-crime.

AI is not the elixir for all forms of security. Although AI-based approaches are becoming more common and cost-effective in most aspects of cybersecurity, they do not provide complete prevention or remediation measures. When a human opponent with an unfaltering stance attacks an intelligent system, there are limits to what an AI can do. It is essential to know that AI is not a factotum and will not be able to handle everything on its own, at least not right now. It actually needs expert human training and supervision to improve over time for the best results. Research shows that artificial intelligence has seemingly positively affected cybersecurity and risks. Hence the continuation of AI and machine learning will take the cybersecurity field to a new level of intelligence.

## References

1. Cybersecurity ventures official annual cybercrime report. https://cybersecurityventures.com/annual-cybercrime-report-2017/ (2022), [Online; accessed 19-May-2022]
2. "global cybersecurity outlook 2022". https://www3.weforum.org/docs/WEF Global Cybersecurity Outlook 2022.pdf (2022), [Online; accessed 19-May-2022]
3. Al-Yaseen, W., Othman, Z., Ahmad Nazri, M.Z.: Multi-level hybrid support vector machine and extreme learning machine based on modified k-means for intrusion detection system. Expert Systems with Applications 67 (01 2017)
4. Banu, R., M, A., C, A., S, A., Ujwala, H., N, H.: Detecting phishing attacks using natural language processing and machine learning. pp. 1210–1214 (05 2019)
5. Baptista, I., Shiaeles, S., Kolokotronis, N.: A novel malware detection sys-tem based on machine learning and binary visualization. pp. 1–6 (05 2019). https://doi.org/10.1109/ICCW.2019.8757060
6. Barbara, D., Couto, J., Jajodia, S., Popyack, L., Wu, N.: Adam: Detecting intrusions by data mining pp. 5–6 (07 2001)
7. Bose, S., Barao, T., Liu, X.: Explaining ai for malware detection: Analysis of mechanisms of malconv. In: 2020 International Joint Conference on Neural Networks (IJCNN). pp. 1–8 (2020)
8. Chowdhury, M., Rahman, A., Islam, M.R.: Malware analysis and detection us-ing data mining and machine learning classification. pp. 266–274 (01 2018). https://doi.org/10.1007/978-3-319-67071-3 33
9. Coull, S., Gardner, C.: Activation analysis of a byte-based deep neural network for malware classification. pp. 21–27 (05 2019). https://doi.org/10.1109/SPW.2019.00017
10. Demetrio, L., Biggio, B., Lagorio, G., Roli, F., Armando, A.: Explaining vulnerabilities of deep learning to adversarial malware binaries (01 2019)

11. Farid, D., Zahidur Rahman, M.: Anomaly network intrusion detection based on improved self adaptive bayesian algorithm. Journal of Computers 5 (01 2010)
12. Feng, F., Zhou, Q., Shen, Z., Xuhui, Y., Lihong, H., Wang, J.: The application of a novel neural network in the detection of phishing websites. Journal of Ambient Intelligence and Humanized Computing (04 2018)
13. Feng, W., Sun, J., Zhang, L., Cao, C., Yang, Q.: A support vector machine based naive bayes algorithm for spam filtering. pp. 1–8 (12 2016)
14. Guan, Z., Li, J., Wu, L.: Achieving efficient and secure data acquisition for cloud-supported internet of things in smart grid. IEEE Internet Things J 4(6), 1934–1944 (09 2017)
15. Hall, M.: Correlation-based feature selection for machine learning. Department of Computer Science 19 (06 2000)
16. Hamamoto, A., Carvalho, L., D. H. Sampaio, L., Abrao, T., Proença, M.: Network anomaly detection system using genetic algorithm and fuzzy logic. Expert Systems with Applications 92 (09 2017)
17. Hamid, Y., Muthukumarasamy, S., Ranganathan, B.: Ids using machine learning - current state of art and future directions. British Journal of Applied Science and Technology 15, 1–22 (03 2016)
18. Hashemi, H., Azmoodeh, A., Hamzeh, A., Hashemi, S.: Graph embedding as a new approach for unknown malware detection. Journal of Computer Virology and Hacking Techniques 13 (08 2017). https://doi.org/10.1007/s11416-016-0278-y
19. Hossain Faruk, M.J., Shahriar, H., Valero, M., Barsha, F., Sobhan, S., Khan, A., Whitman, M., Cuzzocrea, A., Lo, D., Rahman, A., Wu, F.: Malware detection and prevention using artificial intelligence techniques (12 2021). https://doi.org/10.1109/BigData52589.2021.9671434
20. Johns, J.: "representation learning for malware classification". https://www.fireeye.com/content/dam/fireeye-www/blog/pdfs/malware-classification-slides.pdf (2017), [Online; accessed 19-May-2022]
21. Kissell, R.L.: Chapter 2 - algorithmic trading. In: Kissell, R.L. (ed.) Algorithmic Trading Methods (Second Edition), pp. 23–56. Academic Press, second edition edn. (2021). https://doi.org/https://doi.org/10.1016/B978-0-12-815630-8.00002-8, https://www.sciencedirect.com/science/article/pii/B9780128156308000028
22. Kolosnjaji, B., Demontis, A., Biggio, B., Maiorca, D., Giacinto, G., Eckert, C., Roli, F.: Adversarial malware binaries: Evading deep learning for malware detection in executables (2018). https://doi.org/10.48550/ARXIV.1803.04173, https://arxiv.org/abs/1803.04173
23. Kreuk, F., Barak, A., Aviv-Reuven, S., Baruch, M., Pinkas, B., Keshet, J.: Deceiv-ing end-to-end deep learning malware detectors using adversarial examples (2018)
24. Luo, B., Xia, J.: A novel intrusion detection system based on feature generation with visualization strategy. Expert Systems with Applications 41, 4139 4147 (07 2014)
25. Mahajan, R., Siddavatam, I.: Phishing website detection using machine learning algorithms. International Journal of Computer Applications 181, 45–47 (10 2018)
26. McLaughlin, N., Doupé, A., Ahn, G., Martinez-del Rincon, J., Kang, B., Yerima, S., Miller, P., Sezer, S., Safaei, Y., Trickel, E., Zhao, Z.: Deep android malware detection. pp. 301–308 (03 2017). https://doi.org/10.1145/3029806.3029823
27. Molina Valdiviezo, L., Furfaro, A., Malena, G., Parise, A.: A simulation model for the analysis of ddos amplification attacks (03 2015)
28. Obotivere, B., Nwaezeigwe, A.: Cyber security threats on the internet and possible solutions. IJARCCE 9, 92–97 (09 2020)

29. Peng, T., Harris, I., Sawa, Y.: Detecting phishing attacks using natural language processing and machine learning. pp. 300–301 (01 2018)
30. Rokon, M.O.F., Islam, R., Darki, A., Papalexakis, E., Faloutsos, M.: Sourcefinder: Finding malware source-code from publicly available repositories in github (10 2020)
31. Sausalito, C.: Cyberwarfare in the c-suite. https://cybersecurityventures.com/hackerpocalypse-cybercrime-report-2016/ (Nov 13, 2020), [Online; accessed 19-May-2022]
32. Schultz, M., Eskin, E., Zadok, F., Stolfo, S.: Data mining methods for detection of new malicious executables. pp. 38–49 (02 2001)
33. Sharma, S., Challa, R., Sahay, S.: Detection of advanced malware by machine learning techniques (03 2019)
34. Shrestha, P., Maharjan, S., Ramirez-de-la Rosa, G., Sprague, A., Solorio, T., Warner, G.: Using string information for malware family identification. pp. 686–697 (11 2014). https://doi.org/10.1007/978-3-319-12027-0 55
35. Syiemlieh, P., Golden, M., Khongsit, Sharma, U., Sharma, B.: Phishing-an analysis on the types, causes, preventive measuresand case studies in the current situation (01 2015)
36. Taghavi Zargar, S., Joshi, J., Tipper, D.: A survey of defense mechanisms against distributed denial of service (ddos) flooding attacks. IEEE Communications Surveys & Tutorials 15, 2046 – 2069 (11 2013)
37. Tandon, R.: A survey of distributed denial of service attacks and defenses (2020). https://doi.org/10.48550/ARXIV.2008.01345, https://arxiv.org/abs/2008.01345
38. Y. K. Lau, R., S. Y., L., Kwok, R.C.W., Xu, K., Xia, Y., Li, Y.: Text mining and probabilistic language modeling for online review spam detection. vol. 2, pp. 1–30 (12 2011)
39. Ye, Y., Chen, L., Hou, S., Hardy, W., Li, X.: Deepam: a heterogeneous deep learning framework for intelligent malware detection. Knowledge and Information Systems 54, 1–21 (02 2018). https://doi.org/10.1007/s10115-017-1058-9
40. Zhu, H.J., You, Z.H., Zhu, Z., Shi, W.L., Cheng, L.: Droiddet: effective and robust detection of android malware using static analysis along with rotation forest model. Neurocomputing 272, 638–646 (01 2018)